\documentstyle[twocolumn,prl,aps,psfig]{revtex}
\tighten
\begin{document}
\preprint{\today}
\draft
%
%
\title{Singular quasiparticle scattering in the proximity of charge
instabilities}
\author{C. Castellani, C. Di Castro, and M. Grilli}
\address{Dipartimento di Fisica, Universit\`a di Roma ``La Sapienza'',\\
Piazzale A. Moro 2, Roma, Italy 00185}
\maketitle
%
%
\begin{abstract}
We analyze the behavior of the dynamic scattering
amplitude between Fermi liquid quasiparticles at
the Fermi surface in the proximity of a charge instability,
which may occur in the high temperature superconducting
cuprates.
Within the infinite-U Hubbard-Holstein
model in the slave-boson large-N technique
we find that, in the absence of
long-range Coulomb forces the scattering amplitude
is strongly singular at zero momentum transfer
close to the phase separation instability
and it has the same form provided
by gauge-field theories.
In the presence of long-range Coulomb forces
the charge instability occurs at finite
wavevectors and concomitantly the scattering is still
singular but anisotropic. Nevertheless it remains strong
over extended regions of the momentum space. In both
cases we show how normal state properties
are largely affected by this scattering.
\end{abstract}
%
%
%
%
It is generally accepted that the understanding of
the pairing mechanism in high $T_c$
superconductors is related to the understanding
of the anomalous behaviour of the normal phase.

The anomalous properties of the normal phase have been interpreted along
two distinct theoretical lines. One possible explanation is that the
low dimensionality of these highly anisotropic systems and their
correlated nature are at the origin of a breakdown of the
Fermi liquid (FL).In particular the proposal of a Luttinger liquid
formation in two dimension \cite{anderson} has been intensively
investigated \cite{sanseb}.
The alternative aptitude has been to accept the Landau theory of
normal FL's as a suitable starting point. The anomalous properties
would then arise as a consequence of strong scattering processes at low
energy between the quasiparticles. Along this line
magnetic scattering has been considered to be
responsible for both the anomalous  properties of the normal
phase and for the superconducting pairing \cite{pines}.
Strong scattering
may even lead to a complete disruption of the FL phase.
In particular it was proposed that excitonic scattering
could give rise to the so called marginal FL \cite{varma},
and could also provide a pairing mechanism.
Singular scattering can also be provided by gauge fields
\cite{nagaosa}, which
arise by implementing the resonating-valence-bond idea in the
t-J model.

In this letter we want to understand whether phase separation (PS)
or the incommensurate
charge density wave (ICDW) instability are
sources of strong scattering besides the above  mechanisms.
Indeed the complex nature of the phase diagram as a function of
doping and temperature indicates that various energy scales
of different nature (magnetic, excitonic,...) of the same order
of magnitude compete to determine the low-energy physics
and may lead to various instabilities, among which
PS or charge instabilities may play a relevant role.

After PS was shown to be present in the phase
diagram of the t-J model \cite{emery,marder}, we pointed out
that PS commonly occurs in models with short
range interaction \cite{GRCDK1,CCCDGR,DG,RCGBK,CGK,CDG,GC,BTGCD},
 provided the strong local
$e$-$e$ repulsion inhibits the stabilizing
role of the kinetic energy. We therefore stressed
that PS and superconductivity can
be related phenomena irrespective of the nature
of the short-range interaction.
Emery and Kivelson \cite{emerykivelson} suggested that,
although long range Coulomb (LRC)
forces spoil PS as a static thermodynamic
phenomenon, the frustrated tendency towards phase
separation may still be important and give rise
to large amplitude collective density fluctuations.
Approaching the problem within a coarse-grained model,
they suggested that these
fluctuations may be responsible for the anomalous
behaviour of the normal phase and for the superconducting
pairing.

To assess the relevance of PS as a mechanism for anomalous
scattering, we here determine the
dynamical effective scattering interactions
among quasiparticles close to a charge instability,
both in the presence and in the absence of LRC forces.
We carry out this analysis within a microscopic treatment
of the Hubbard-Holstein model in the infinite-U limit.
We find that, both in the presence and in the absence of
LRC forces, the dynamic effective interaction
turns out to have a  singular behaviour, strongly affecting
the single-particle and the transport scattering time.

-{\it {The model}} - Although
our results are quite generic of models with PS,
to be specific we use as a simple paradigm
the two-dimensional Hubbard model with an
additional dispersionless phonon mode $A$ coupled {\it \`a la } Holstein
\begin{eqnarray}
 H & = & -t \sum_{\langle i,j \rangle , \sigma}
\left( c^\dagger_{i\sigma} c_{j\sigma} + H.c.\right)
-t' \sum_{\langle \langle i,j \rangle \rangle , \sigma}
\left( c^\dagger_{i\sigma} c_{j\sigma} + H.c.\right) \nonumber \\
+ & \omega_0 & \sum_i A^\dagger_i A_i + g \sum_{i,\sigma}
\left( A^\dagger_i+A_i\right) \left( n_{i\sigma} -\langle n_{i\sigma} \rangle
\right) \nonumber \\
- & \mu_0 & \sum_{i\sigma} n_{i\sigma}
+ U\sum_i n_{i\uparrow}n_{i\downarrow} +
{\sum_q}' {V_C \over \sqrt{G^2(q) -1}}
\rho_q \rho_{-q}
,\label{HHHam}
\end{eqnarray}
where $\langle i,j \rangle$ and $\langle \langle i,j \rangle\rangle$
indicate nearest-neighbor and next-nearest-neighbor sites respectively
and $\sum_\sigma n_{i\sigma}=
\sum_\sigma c^\dagger_{i\sigma} c_{i\sigma} $ is the local
electron density, which in momentum space is given by
$\rho_q \equiv \sum_{k,\sigma} c^\dagger_{k+q,\sigma}c_{k,\sigma}$.
The last term of Eq.(1)
is a Coulombic potential between electrons on a
two-dimensional square lattice (with lattice
spacing $a$ in the x and y directions) embedded in a three-dimensional space
with a separation $d$ between the planes in the z direction.
The dielectric constants in the plane and perpendicular to it
are $\epsilon_{\parallel} $ and $\epsilon_{\perp} $ respectively
and the Coulombic coupling constant $V_C=
e^2 d /2 \epsilon_{\perp} a^2 $.
The momentum dependence of the potential on the z=0 plane
is found to be $G(q_x,q_y) = {\epsilon_{\parallel}
\over \epsilon_{\perp} (a/d)^2 } \left(
\cos (aq_x) + \cos (aq_y) -2 \right) -1$.
As usual, the sum in the Coulombic potential
does not include the zero-momentum component, since
we are supposing that the diverging $q=0$ interaction between the
electrons is canceled by the contribution of a uniform positively
charged ionic background.

Since we are interested in strong local repulsion
we take the limit $U\to \infty$, which gives rise to the local
constraint of no double occupancy $\sum_\sigma n_{i\sigma} \le 1$.
To implement this constraint we use a standard slave-boson technique
\cite{barnes}, by performing the usual substitution
$c^{\dagger}_{i \sigma}\rightarrow c^{\dagger}_{i
\sigma}b_i, \,\,\, {c}_{i \sigma}\rightarrow b^{\dagger}_i c_{i
\sigma}$ and introducing  a Lagrange multiplier field $\lambda_i$.
The quartic Coulombic term can be decoupled by
a Hubbard-Stratonovich transformation introducing
an additional real bosonic field $Y_i$.
Within the large-$N$ expansion we assume that the spin index
runs from 1 to $N$ and we relax the constraint to the form
$\sum_\sigma c^{\dagger}_{i\sigma} c_{i\sigma} +b^{\dagger}_ib_i =
{N \over 2}$. A suitable rescaling of the hoppings
$t \rightarrow t/N$ and $t' \rightarrow t'/N$ must, in this
model, be joined by the similar rescaling of the $e$-$ph$
coupling $g \rightarrow {g/ {\sqrt{N}}}$
and of the Coulomb interaction $V_C \to V_C/N$ in order to
compensate for the presence of $N$ fermionic degrees of freedom.

The model can first be solved in the mean field ($N=\infty$)
approximation by setting the $b_i$ and $\lambda_i$ bosons
to their constant self-consistent values
$b_0$ and $\lambda_0$ respectively.
The system then results into free quasiparticles with a shifted
chemical potential $\mu=\mu_0-\lambda_0$ and a dispersion
$E_k= -2t r_0^2 \varepsilon_k $ with $\varepsilon_k \equiv \left(\cos (ak_x)
+\cos (ak_y) \right)
+ (t'/t) \left(\cos (ak_x+ak_y) + \cos (ak_x-ak_y) \right)$
where $r_0^2=b_0^2/N=\delta /2$.
For any finite doping $\delta$ at T=0
the system is a Fermi-liquid where the  mean-field value of the
slave-boson field $b_0$ multiplicatively reduces the
hoppings, $t \to tb_0^2$, $t' \to t'b_0^2$,
thus enhancing the effective mass
of the quasiparticles. At this
level  the mean-field self-energy does not introduce a finite
quasiparticle lifetime.

The effective interaction leading to scattering between quasiparticles
arises from the exchange of the bosonic fields
in the $1/N$ corrections beyond mean field approximation.
One can define a four-component field  $\alpha^{\mu}=(\delta
r,\,\,\, \delta \lambda, \,\,\, \phi, \,\,\, Y)$.
$\phi_i=(A_i^{\dag}+A_i)/(2\sqrt{N})$ is the lattice displacement field,
and $\delta r_i$ and $\delta \lambda_i$
are the fluctuating part of the $b_i$-field amplitude and of the
the Lagrange multiplier respectively.
The leading-order
expressions of the effective scattering amplitude  in the
particle-hole channel can be written as
\begin{equation}
\Gamma (k,k';q,\omega)= -\sum_{\mu \nu}
{\Lambda}^{\mu}  \left(k', -q
\right) D^{\mu \nu} \left( q, \omega \right)
{\Lambda}^{\nu} \left(k, q \right). \label{gamma}
\end{equation}
where $\Lambda^\mu$ are the vertices
coupling the fermionic quasiparticles to the bosons,
$\Lambda_r(k,q)     =  -2tr_0^2 \left(\varepsilon_{k+q/2}
+\varepsilon_{k-q/2} \right) $, $\Lambda_\lambda (k,q)= i  $,
$\Lambda_\phi (k,q) = -2g$ and $\Lambda_Y(k,q)=i$.
$D^{\mu \nu}(q,\omega)  =
\langle \alpha^\mu(q,\omega)\alpha^\nu(-q,-\omega)\rangle
= N^{-1}(2B+
\Pi(q,\omega))^{-1}_{\mu \nu}$ is the leading order boson propagator
with self-energy corrections given by the fermionic
bubbles $\Pi^{\mu \nu}$ (which include
quasiparticle-boson vertices). $B$ is the bare boson-propagator
matrix.
An expression similar to Eq. (\ref{gamma})
holds for the scattering amplitude in the particle-particle channel.

- {\it The results} -
The evaluation of the density correlation function
$P(q,\omega) \equiv (1/N)
\sum_{\sigma \sigma'}\langle n_{\sigma}(q,\omega)
n_{\sigma'}(-q,-\omega)\rangle$
provides information
on the stability of the system.  In particular
a divergence in the static density-density correlation function
$P(q,\omega=0)$ signals the  occurrence of
PS (at $q \to 0$)
or of CDW instabilities (at finite $q$'s).
A complete investigation of the static and dynamical properties of the
present model together with the analysis of its stability
was  already carried out in a previous work \cite{BTGCD}.
Here we just mention that, within the
present formalism, the model  (\ref{HHHam})
displays a phonon-driven
charge instability even for rather small e-ph coupling.
In the absence of LRC forces  this instability occurs
before any other finite $q$ instability up
to intermediate-large doping. At large
doping the instability requires larger e-ph couplings
and occurs at finite $q\approx 2k_F$ in some directions
[in particular (1,0) and (0,1)]
signalling the occurrence of incommensurate CDW. The introduction
of LRC forces eliminates the small-$q$ divergence in the
static correlation function
always giving rise to finite $q$ instabilities. The critical $q$,
in this case, is not related to any pseudonesting
of the Fermi surface but it depends on the strength of
the LRC forces and on the momentum dependence of the
poles in the divergent static correlation function with
only short-range forces. Pair formation is always found near
the instabilities.

As already announced, we want to report on the
behavior of the quasiparticle scattering
close to both the PS and
the CDW instability. It is worth noting that
a divergent scattering amplitude will follow from
a divergent correlation function $P$.
Indeed a divergent boson propagator
enters in the expression of both quantities
establishing a clear connection between the charge instability
and the singular quasiparticle scattering.

In this regard we carried out an extensive analysis
of the real and imaginary parts of  the scattering amplitude between the
quasiparticles on the Fermi surface.

As expected, near the PS instability ($V_C=0$), the anomalous behavior
of $\Gamma$ is identified to be of the form
\begin{equation}
\Gamma (q,\omega) \approx
- {1 \over B
q^2 - i\omega {C\over q} +D} \label{fitgamsr}
\end{equation}
As shown in Fig. 1a, for the model with $t=0.5$eV, $t'/t=-1/6$,
$\omega_0=0.04$eV and $g=0.194$eV,
$D=D(\delta-\delta_c)$ vanishes linearly
when, for a given $g$, the instability takes place
at the critical doping $\delta_c=\delta_c(g)$.
For the same values of the parameters,
Fig.1b reports the real static scattering amplitude
$\Gamma(q,\omega=0)$ as a function of transferred momentum $q$
at various dopings close to the $q=0$ instability.
In this case the singular part of the scattering amplitude displays a quite
isotropic behavior (at $q_c=0$).

In the absence of LRC forces the singular behavior
of $\Gamma(q \to 0,\omega=0)$ at the PS instability is
by no means surprising within a FL framework. Indeed, the
FL expression for the compressibility is $\kappa=2\nu^*/\left(1+2\nu^*
\Gamma_\omega \right)=2\nu^*(1-2\nu^*\Gamma_q)$, where $\Gamma_\omega$
and $\Gamma_q$ are the standard dynamic ($\omega \to 0, q=0$) and
static ($q\to 0, \omega=0$) limit of the scattering amplitude.
This indicates that a divergent $\kappa$,
when the quasiparticle mass remains
finite ($\nu^*<\infty$), only happens when
$F_0^s\equiv 2\nu^*\Gamma_\omega \to -1$ (Pomeranchuk criterion).
At the same time if $\Gamma_q \to \infty$.
 We like to point out here that the above arguments keep their full
validity irrespective of the mechanism leading to PS.

In the presence of LRC forces the singular part of $\Gamma$
can be written as
\begin{equation}
\Gamma ({\mbox{\boldmath $q$}},\omega) \approx
- {A \over \omega_{\mbox{\boldmath $q$}} - i\omega} \label{fitgamlr}
\end{equation}
where
$\omega_{\mbox{\boldmath $q$}} = D'+B'\vert {\mbox{\boldmath $q$}}
-{\mbox{\boldmath $q$}_c} \vert^2$
The behavior of the mass term $D'$ is shown in Fig. 2a
for the model with $t=0.5$eV, $t'/t=-1/6$, $V_C=1.1$eV, $\omega_0=0.04$eV and
$g=0.240$eV as a function of $\delta-\delta_c$.
For these parameters the instability first occurs at $\delta_c=0.194$
and ${\mbox{\boldmath $q$}}_c\approx(\pm 0.28/a,\pm 0.86/a)$, or
${\mbox{\boldmath $q$}}_c\approx(\pm 0.86/a,\pm 0.28/a)$.
Analogously to Fig. 1b, Fig. 2b displays the strong
doping dependence of the static scattering
amplitude as a function of momenta in the
${\mbox{\boldmath $q$}}_c$ direction. However, as shown in Fig. 3,
the scattering is
quite strong, although non-singular, in all directions
for $\vert {\mbox{\boldmath $q$}} \vert \approx \vert
{\mbox{\boldmath $q$}}_c \vert$.
We also checked that the (almost) isotropic
contribution to the static scattering amplitude it is much less
fragile under doping variations.

The imaginary term in the denominators in the
r.h.s. of Eqs.(\ref{fitgamsr}) and (\ref{fitgamlr})
reproduces on a wide range of
transferred momenta $q$ the behaviour of the imaginary
part of the mean field fermionic
polarization bubble $Im\left[ P^0({\mbox{\boldmath $q$}},
\omega)\right] \propto
\omega /q$ at small $\omega $.  This indicates, that,
despite the complicated formal structure of the
scattering amplitude (\ref{gamma}) arising from the
matrix form of the boson propagator $D^{\mu \nu}$,
near the instability a simple RPA-like structure results
in the final expression. The forms (\ref{fitgamsr})
and (\ref{fitgamlr}) are, therefore,
generic of PS or charge instabilities.

 It is apparent that the effective interaction
(\ref{fitgamsr}) in the isotropic pure-short-range case,
has the same form
as the one mediated by gauge fields \cite{nagaosa}, if it were
not for the doping-dependent mass term and for the completely
different origin of the singularity.
The LRC forces make the scattering anisotropic
rendering it more similar to the
magnetic fluctuation case \cite{pines}, even though the
really singular behaviour is for differently oriented momenta
and large scattering is present in any direction.

Near PS at $\delta=\delta_c$, the electrons are coupled
to a mode, which is soft at long wavelength, and the same results
of Ref.\cite{nagaosa} apply. Therefore in two dimensions
the inverse scattering time is proportional to $T^{2/3}$
and the resistivity $\rho$ is proportional to $T^{4/3}$ with a
crossover temperature $T^*$ to the standard FL behavior,
which in our case turns out to be $T^* \propto (\delta -\delta_c)^{3/2}$.
The resistivity is anomalous but it is not linear in T as in the
optimally doped high $T_c$ materials. An additional
the limitation of the pure short-range case may be
the difficulty in approaching the instability line which might
be embedded in a region stabilized by the Maxwell construction.

In the more physical case including LRC forces, the proposal
of associating the anomalous
properties of the normal phase to the presence of a $T=0$
quantum critical point \cite{pinesQCP,varmaprep} has
a natural realization within
the charge-instability mechanism.
The outcome for the transport properties should be similar
to the one described within the nearly antiferromagnetic
FL scenario \cite{pines}, i.e. a linear resistivity for temperatures
larger than a crossover temperature $T^*$, which, in our case would
be proportional to $D'$ in Eq.(\ref{fitgamlr}).
However, this result was questioned in Ref.\cite{hlubina}
for the AF fluctuations because only few ``hot'' points
on the Fermi surface are connected by singular interactions,
with an anomalous inverse scattering time proportional to
$\sqrt{T}$. Generically the resistivity would be dominated
by an inverse scattering time proportional to $T^2$, then
linear resistivity would only appear for $T \gg T^*$. In our
case, instead, a strong rather isotropic scattering persists
rendering the above limitation less effective.

The above scenario connecting charge instabilities to the
anomalous normal properties of the superconducting cuprates
raises the problem of the relevance of charge fluctuations
in these systems \cite{erice}. Indeed commensurate CDW
have been recently observed \cite{tranquada} in a related
compound (${\mathrm {La}}_{2-x-y} {\mathrm {Nd}}_y {\mathrm {Sr}}_x
{\mathrm {CuO}}_4$ with $x=0.12$ and $y=0.4$)
possibly due to the pinning of electronic dynamical CDW
by a suitable underlying lattice structure. It is therefore
quite important to assess the electronic
origin of this CDW, e.g. by detecting non-linear effects like
electric-field-induced depinning. This would
strongly support the existence of dynamical CDW fluctuations
when pinning is not effective leaving the materials in
the metallic phase.

%
%
%
%

%
%
\begin{figure}
\centerline{\hbox{\psfig{figure=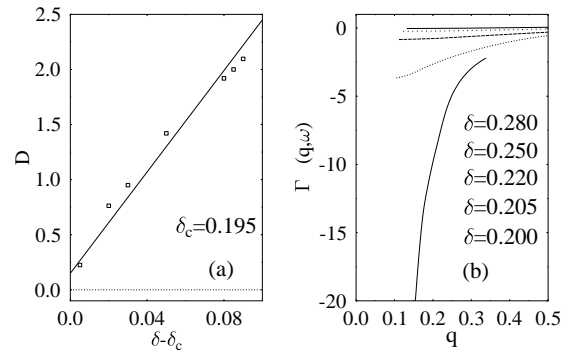,width=5.5cm,angle=-90}}}
\caption{(a) Mass $D$ as a function $\delta-\delta_c$
for $t=0.5$eV $t'=-1/6t$, $V_C=0$, $\omega_0=0.04$eV and
$g=0.194$eV. (b)
Static scattering amplitude for the same parameters as in (a)
as a function of the transferred
momentum ${\mbox{\boldmath $q$}}$ in the (1,0) direction. The doping
 $\delta=0.2,0.205, 0.22,0.25, 0.28$ increases
from the lower solid line to the upper solid line.}
\label{FIG1}
\end{figure}
\begin{figure}
\centerline{\hbox{\psfig{figure=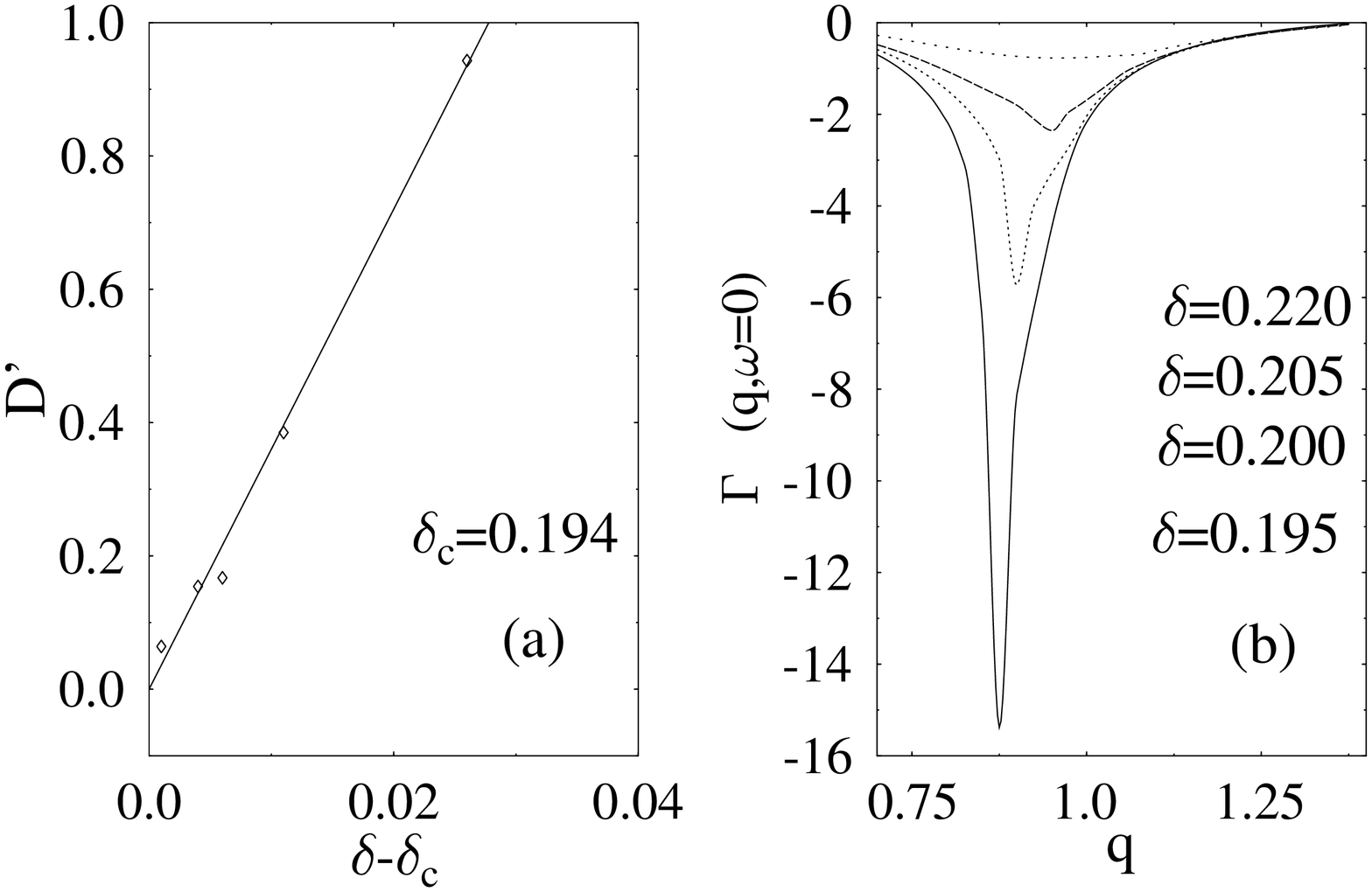,width=5.5cm,angle=-90}}}
\caption{(a) Mass $D'$ as a function $\delta-\delta_c$
for $t=0.5$eV $t'=-1/6t$, $V_C=1.1$eV, $\omega_0=0.04$eV and
$g=0.312$eV. (b) Static scattering amplitude for the same parameters as in (a)
as a function of the transferred
momentum ${\mbox{\boldmath $q$}}$ in the ${\mbox{\boldmath $q$}}_c
\approx(\pm 0.28/a,\pm 0.86/a)$, direction.
The doping $\delta=0.195,0.2, 0.205, 0.22$ increases
from the lower solid line to the upper solid line.}
\label{FIG2}
\end{figure}
\begin{figure}
\centerline{\hbox{\psfig{figure=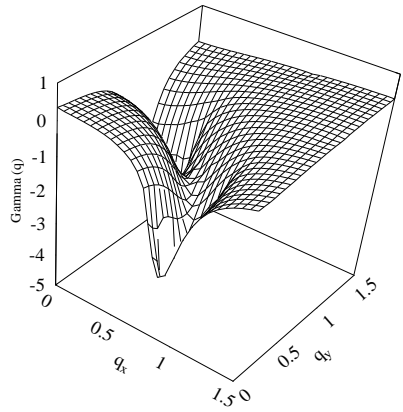,width=7.5cm,angle=-90}}}
\caption{Momentum dependence of the static
scattering amplitude for the same parameters as in Fig. 2a
at $\delta=0.195$.}
\label{FIG3}
\end{figure}


\begin{references}
\bibitem{nota}With this value for $V_C$ electrons on neighboring
cells feel a repulsion of $\sim 0.2$eV when $\epsilon_\parallel=5$
and $\epsilon_{\perp}=25$ and $d=13$A. These values are
rather typical for ${\mathrm {La_{2-x}Sr_xCuO_4}}$ compounds.
\end{references}
\end{document}